# Buccal-Lingual Analysis and Inflammation Tracking in Oral Soft Tissues Using Quantitative Ultrasound: A Preclinical Study

**Daria Poul**
*Dept. of Radiology*
*University of Michigan*
*Ann Arbor, MI, USA.*
ssheykho@umich.edu

**Amanda Rodriguez**
*Dept. of Periodontics, College of Dentistry, University of Illinois Chicago, Chicago, IL, USA.* arodr368@uic.edu

**Ankita Samal**
*Dept. of Periodontics, University of Iowa College of Dentistry, Iowa City, IA, USA.*
ankita-samal@uiowa.edu

**Carole Quesada**
*Dept. of Radiology*
*University of Michigan*
*Ann Arbor, MI, USA*
cquesada@umich.edu

**Ted Lynch**
*Sun Nuclear Corporation, Norfolk, VA, USA.*
tlynch@mirion.com

**Cristel Baiu**
*Dept. of Medical Physics, University of Wisconsin-Madison, Madison, WI, USA.*
baiu@wisc.edu

**J. Brian Fowlkes**
*Dept. of Radiology and Dept. of Biomedical Engineering, University of Michigan Ann Arbor, MI, USA.*
fowlkes@umich.edu

**Hsun-Liang Chan**
*Division of Periodontology, College of Dentistry, The Ohio State University, Columbus, OH, USA* chan.1069@osu.edu

**Oliver D. Kripfgans**
*Dept. of Radiology and Dept. of Biomedical Engineering, University of Michigan Ann Arbor, MI, USA.*
greentom@umich.edu

***Abstract*—Four out of 10 adults aged 30 years or older in the USA are impacted by periodontal (gum) diseases which span a spectrum of inflammatory conditions. Currently, a subjective, invasive and a semi-quantitative approach, termed bleeding on probing, is employed in clinics for inflammation assessment. The long-term goal of this study is to fill the current clinical diagnostic gap in dentistry by proposing quantitative ultrasound (QUS)-based biomarkers for inflammation diagnosis and monitoring. Here, as one of the early works in this area, we investigated two QUS parameters for characterizing periodontal inflammation in gingival tissues using a longitudinal preclinical porcine study. These are ultrasound attenuation coefficient slope (ACS) and backscatter intensity (BSI). Our preclinical study included eight pigs imaged intraorally (frequency: 24 MHz) at four bi-weekly timepoints from week 0 (healthy) to week 6 (post inflammation inoculation) at their interproximal (mesial) sites of the third premolars (PM3-Mes) from all four quadrants. Moreover, we compared gingival tissues surrounding a tooth at lingual/palatal side versus buccal (cheek) side at the second molar (M2-Dis) oral sites in healthy condition. Our results showed that gingival ACS at all inflammation inoculation timepoints were significantly lower than healthy gingival ACS (mean: 1.69; SD: 0.53 dB/cm.MHz) using a mixed effect analysis (week 6: mean: 1.08; SD: 0.25 dB/cm.MHz). Longitudinal comparison of BSI did not demonstrate any statistical significance at this oral site. Moreover, we showed that gingival ACS and BSI at lingual/palatal versus buccal sides of M2-Dis did not exhibit any statistical significance. These ACSs were linearly correlated with R-squared = 0.89 and a correlation slope of 0.87. For BSI, Bland-Altman analysis demonstrated that there is no systemic difference in mean BSI, although variability was high. These findings highlight the promising potential of intraoral ultrasonography paired with QUS to complement current standard of care in dentistry.***

## I. Introduction

Over 46% of adults aged 30 years or older in the USA are affected by periodontal diseases, a spectrum of gum inflammatory conditions. Associated with pain and systemic diseases [1], periodontal diseases negatively affect quality of everyday life and impose financial burdens on patients and the healthcare system. In dentistry, two of the major approaches routinely used in clinics for assessment of periodontal diseases or monitoring gum health are probing and radiographic imaging. In periodontal probing, a marked probe is inserted between gingiva (gum) and tooth, and the pocket depth and bleeding are read to assess inflammation associated with clinical sign of gingival recession and attachment loss. This method is subjective, late-stage indicator and also uncomfortable for patients. Radiographic imaging provides clinicians with a diagnostic lens for bone loss assessment associated with inflammation; however, they expose patients with ionizing radiation, are qualitative while also not offering high contrast for imaging soft tissues. Due to limitations of these approaches in current clinical standard of care [2], research into applications of ultrasonography in dentistry has been emerging [3, 4]. These studies include using ultrasound Bmode for image-based texture and intensity analysis [5, 6] and also landmark measurements of periodontal tissues with clinical meaning such as soft tissue thicknesses [7-10]. Ultrasonography has also been investigated for blood flow quantification of periodontal tissues with pathological conditions [11, 12]. In our recent studies, we investigated the first implementation of quantitative ultrasound (QUS) techniques in periodontology. This includes ultrasound speckle statistical modeling to characterize gingival versus mucosal oral tissues [13, 14] and the quantification of gingival ultrasound attenuation using validated standard techniques [15, 16]. In this longitudinal, preclinical study, we continue that line of investigations by addressing two questions:

1) *How do gingival QUS signatures change in response to periodontal inflammation inoculation (premolar 3- mesial)*;

2) *Are QUS signatures of gingival tissues measured at the buccal (cheek) side different than the lingual-palatal side for tissues surrounding a specific tooth (molar 2-distal).*
Here, we aimed to complement our recent work [15, 16] by extending the longitudinal inflammation analysis to a new oral site (PM3-Mes) as well as by conducting the new buccal-lingual investigation using QUS.

## II. Theory

Here, we focused on two QUS parameters of ultrasound attenuation coefficient slope (ACS) and backscatter intensity (BSI). ACS was estimated using the spectral difference method based on a reference phantom with known attenuation coefficient (AC), $\alpha_r(f)$, scanned under a similar imaging setup [17]. For details of the attenuation estimation implementation and associated validation of the technique, we refer to our recently published study [15]. To summarize, we estimated the tissue AC, $\alpha_s(f)$, by taking the ratio of the measured ultrasound echo signals of tissue to the reference phantom, denoted by $RS(f,z)$ and using the spectral-domain relationship below.

$$Ln\,[RS(f,z)] = -4\big(\alpha_s(f) - \alpha_r(f)\big)(z - z_0) + Const. \quad (1)$$

AC as a function of frequency was estimated over a useable frequency range by using the time-gated signals and employing linear fitting along depth for each spectral component. Here, $z$ is the imaging depth, $z_0$ and $f$ denote phantom start depth and ultrasound frequency, respectively. ACS, $\beta$, is estimated using a one-parameter linear frequency modeling of AC. The units of AC and ACS are dB/cm and dB/cm.MHz, respectively. Backscatter intensity was estimated using the magnitude of the echo envelope.

## III. Method

Our preclinical cohort included eight pigs; each scanned at all four oral quadrants of mandibular (MAND) and maxillary (MAX) lefts and rights. Four longitudinal bi-weekly timepoints were considered: week 0 (healthy), week 2, 4, and 6 (inflammation). Ultrasonography was performed at the frequency of 24 MHz intraorally using an imaging system (ZS3) paired with a toothbrush-size transducer (L30-8) both from the Mindray Innovation Center, NA, San Jose, CA, USA. IQ imaging data were exported for QUS analysis, and the Bmode images were reconstructed by envelope detection using a Hilbert transform followed by logarithmic compression. All postprocessing was performed in MATLAB (MathWorks Inc., Natick, MA, USA).
Two oral sites were enrolled from all pigs: (a) premolar 3-mesial (PM3-Mes) for longitudinal inflammation tracking at all four timepoints, (b) molar 2- distal (M2-Dis) for buccal versus lingual (lower jaw)/palatal (upper jaw) analysis at healthy condition, i.e. week 0. In Figure 1, (a) demonstrate transducer orientation for lingual versus buccal intraoral imaging. In (b) and (c) on this figure, two representatives Bmode images of porcine periodontal tissues at M2-Dis are presented from its lingual and buccal sites, respectively. Important anatomical structures are also annotated.

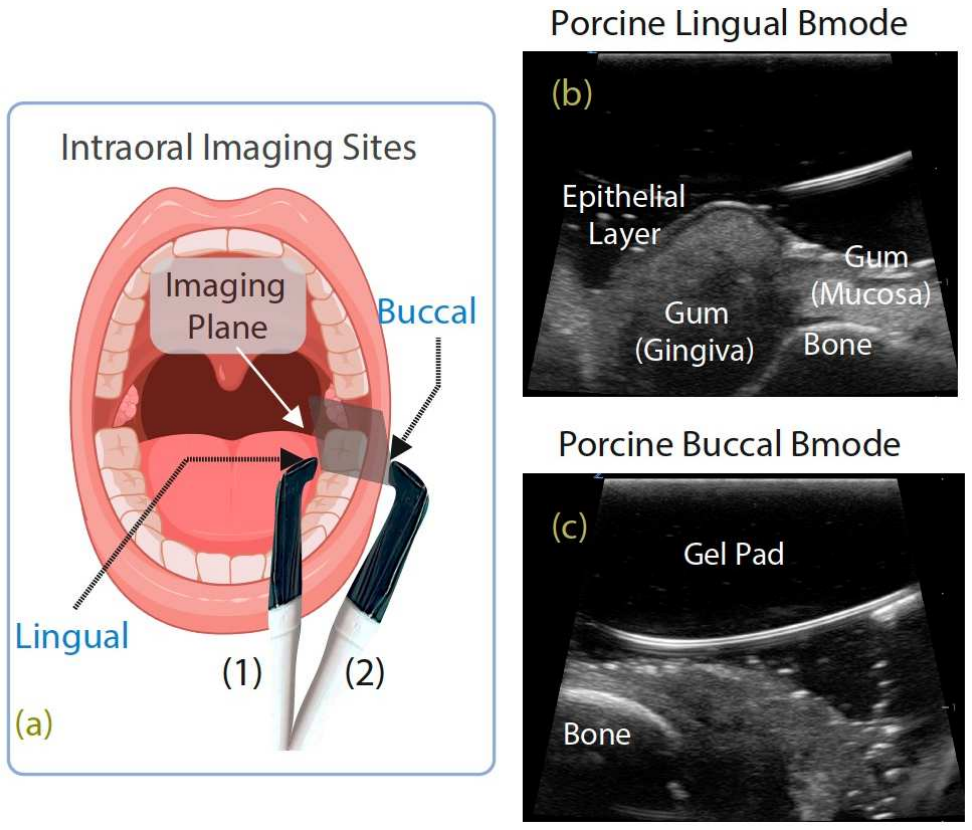

Figure 1: (a) Illustration of lingual and buccal oral sites, the transducer placement for intraoral imaging of (1) lingual and (2) buccal sides. The imaging plane for associated intraoral ultrasonography is also shown. Two porcine Bmode images from (b) lingual and (c) buccal sides of M2-Dis are shown. Important anatomical landmarks and imaging gel pad are annotated.

Inflammation was induced at PM3-Mes using two complementary methods: 1) placing suture around tooh to increase bacterial accumulation at the gingival pocket, 2) a weekly injection of periodontitis- inducing agent into the gingival tissues (Porphyromonasgingivalis, Fusobacterium nucleatum, and Treponemadenticola; 5 $\mu L/min$; one minute). For statistical comparison of ACS at inflammation timepoints versus week 0, a linear mixed effect model was employed (equation (2)) with week included as the fixed effect and the animal-quadrant combination was included as a random effect to account for repeated longitudinal measurements from the same animal-quadrant site at four bi-weekly timepoints. This model assigns a random intercept to each specific animal-quadrant combination, such as Animal2-(MAND/L) between timepoints. For lingual/palatal versus buccal comparison, statistical significance was tested using a two-tailed paired t-test.

$$ACS \sim Week + (1|AnimalID{:}\,Quadrant) \quad (2)$$

## IV. Results and Discussion

### A. ACS-based Inflammation Characterization

Figure 2 (a) shows an example of Bmode image of a PM3 at the inflammation timepoints of week 2 from the left mandibular quadrant with an ROI selected within gingiva for ACS estimation. In panel (b) of this figure, linear frequency modeling of AC as a function of frequency is presented with the estimated ACS of 1.25 dB/cm.MHz reported.

Figure 3 and Figure 4 summarize ACS and BSI estimations of the pig cohort, respectively, measured at their PM3-Mes oral sites comparing healthy versus the three inflammation timepoints. The corresponding p-values for QUS comparisons of healthy and inflammation timepoints are reported in Table 1. We note that a few scans were excluded from the analysis due to the presence of imaging artifacts at the gingival regions, or if the available region for placement of a region of interest (ROI) for ACS estimation was too small. These results show that ACS values at three inflammation timepoints were lower than the healthy condition, with p<0.05. Mean gingival ACS at week 0 was $1.69 \pm 0.53$ dB/cm.MHz, while for weeks 2, 4, and 6 were $1.18 \pm 0.35$ dB/cm.MHz, $1.03 \pm 0.47$ dB/cm.MHz, $1.08 \pm 0.25$ dB/cm.MHz, in that order. BSI estimations did not show any statistical significance between healthy and inflammation timepoints (p>0.05). We hypothesize that the decrease in ACSs at the inflammation timepoints compared with healthy condition is associated with lower ACS of water from increased accumulation of water content in inflammation (edema) within the gingival tissues. Periodontal tissues contain various fibrous structures that provide mechanical support to gingiva (gum), likely contributing to their heterogeneous composition and higher variations in measured ACS. These findings suggest that ACS has a promising potential as a QUS parameter for characterizing periodontal inflammation.

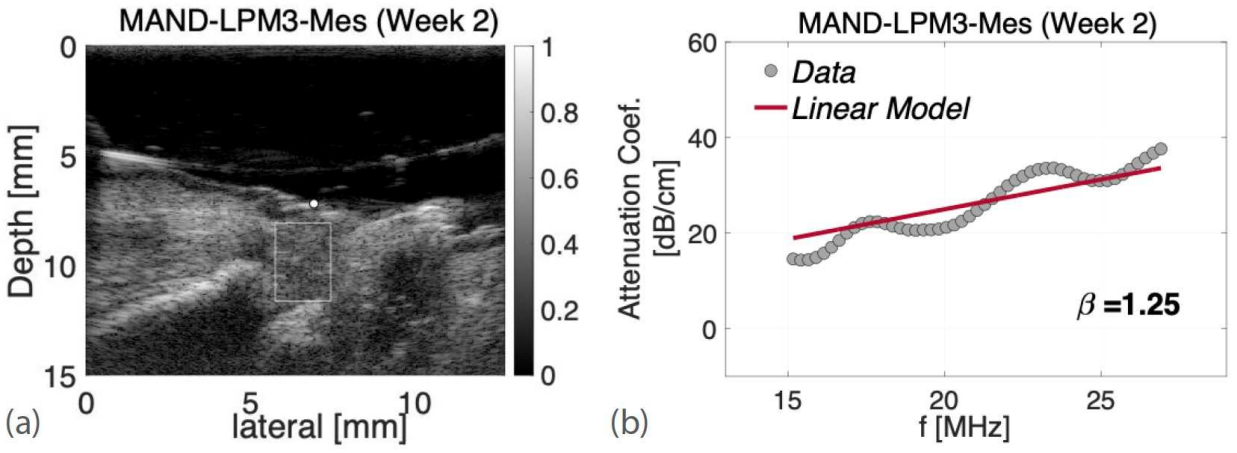

Figure 2. (a) Bmode image and ROI placement for ACS estimation of MAND-LPM3 at week 2, (b) associated linear modeling of the AC as a function of frequency with estimated ACS reported on the bottom right.

### B. *Buccal vs. Lingual-palatal QUS Analysis*

Periodontal tissues surrounding a tooth differ at their buccal side versus lingual/palatal side in terms of some of their anatomical signatures such as landmark tissue thicknesses. To demonstrate such differences, we have presented a histology image of a premolar tooth of our pig cohort in Figure 5, which was stained using the Masson's Trichrome technique [13].

Here, lingual and buccal sides along with three periodontal landmarks of bone, enamel (tooth), and alveolar bone crests are shown. On the lingual and buccal sides, we have annotated two landmark distances that show tissue thicknesses at two sites: $d_1$ denotes the perpendicular distance between the alveolar bone crest and the surface of gingival tissues, i.e. epithelial layer and $d_2$ is the distance between the gingival attachment point to enamel and epithelial layer. As reported in literature [18] and demonstrated in Figure 5, periodontal tissues at lingual sides are generally thinner than buccal sides. Thus, we investigated variations of QUS parameters measured at these two sites.

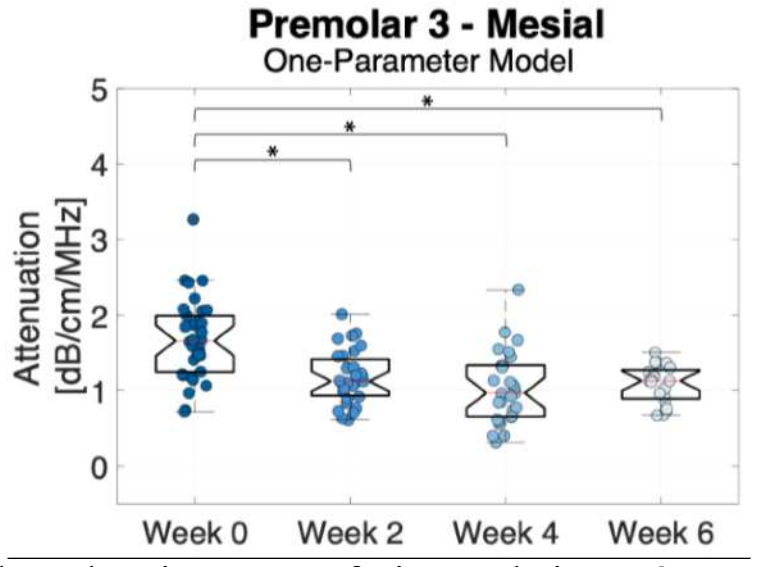

Figure 3: Boxplots showing ACS of pigs at their PM3-Mes sites across all timepoints. Asterisks show statistical significance (p<0.05). See Table 1.

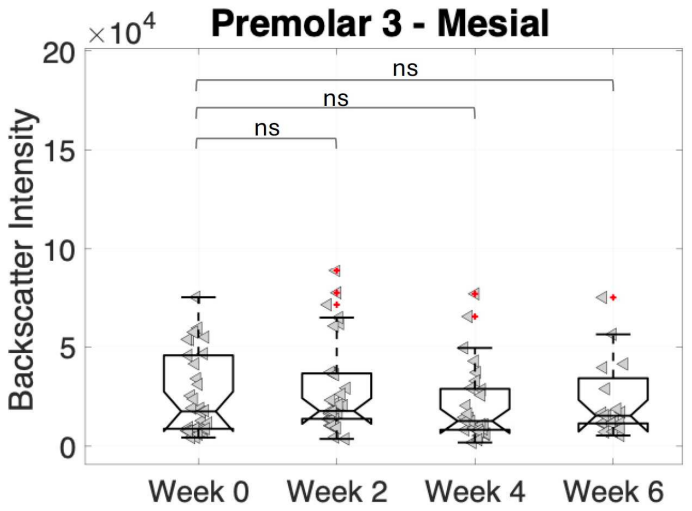

Figure 4: Boxplots showing BSI of pigs at their PM3-Mes sites across all timepoints. ns denotes no statistical significance (p>0.05). See Table 1.

Table 1: P-values for longitudinal comparison of gingival ACS and BSI at healthy versus three inflammation timepoints using the mixed effect model.

| Week 0 vs. | Week 2 | Week 4 | Week 6 |
|---|---|---|---|
| ACS *p-value* | 1.2e-06 | 3.0e-09 | 8.6e-07 |
| BSI *p-value* | 0.97 | 0.1175 | 0.79194 |

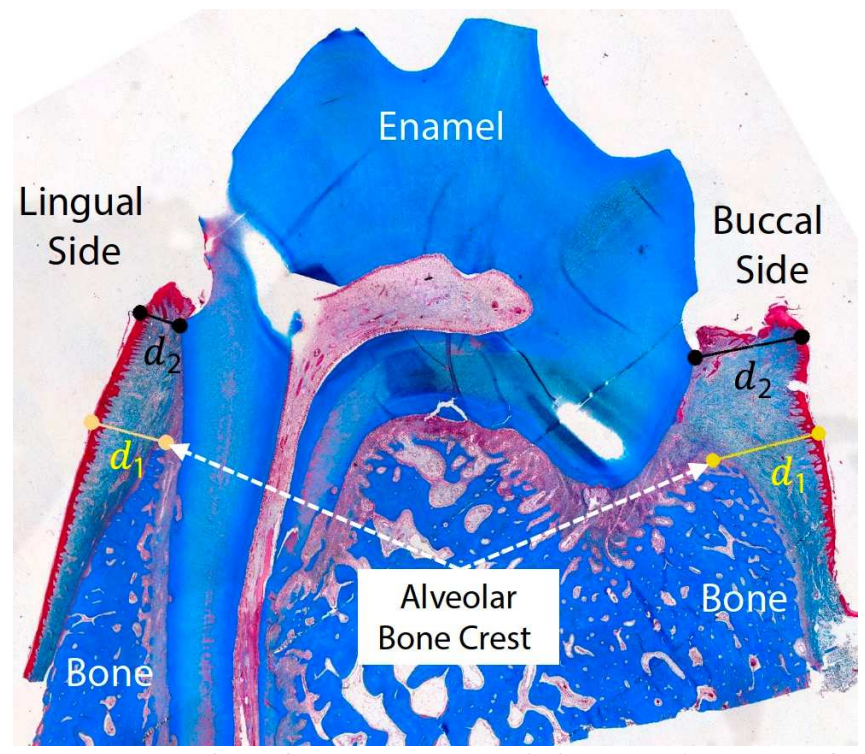

Figure 5: (a) A sample of Masson's Trichrome image of a premolar tooth comparing lingual and buccal periodontal tissues and two landmark anatomical thicknesses of $d_1$ and $d_2$ at both sides.

In Figure 6, we have shown (a) ACS boxplots and (c) BSI boxplots, comparing buccal and lingual/palatal sides of M2-Dis at week 0. Paired student's *t-tests* of mean values exhibited no statistical significance in either ACS or BSI between the two sides (p>0.05). The mean (standard deviation) ACS was $1.29 \pm 0.38$ dB/cm.MHz for lingual/palatal side and $1.19 \pm 0.35$ dB/cm.MHz for the buccal side. We further examined ACS relationship in Figure 6 (b), using a zero-intercept linear regression model for the two sides. The fitted slope was 0.87, with a 95% confidence interval of 0.74 - 1.00, indicating a reasonably close correlation between ACS values measured at the two sides, with interval extending to one. The $R^2$ value of 0.89 indicates a reasonable goodness of the linear fit. These findings suggest that the two sides of the gingiva do not show distinct ACS signatures.

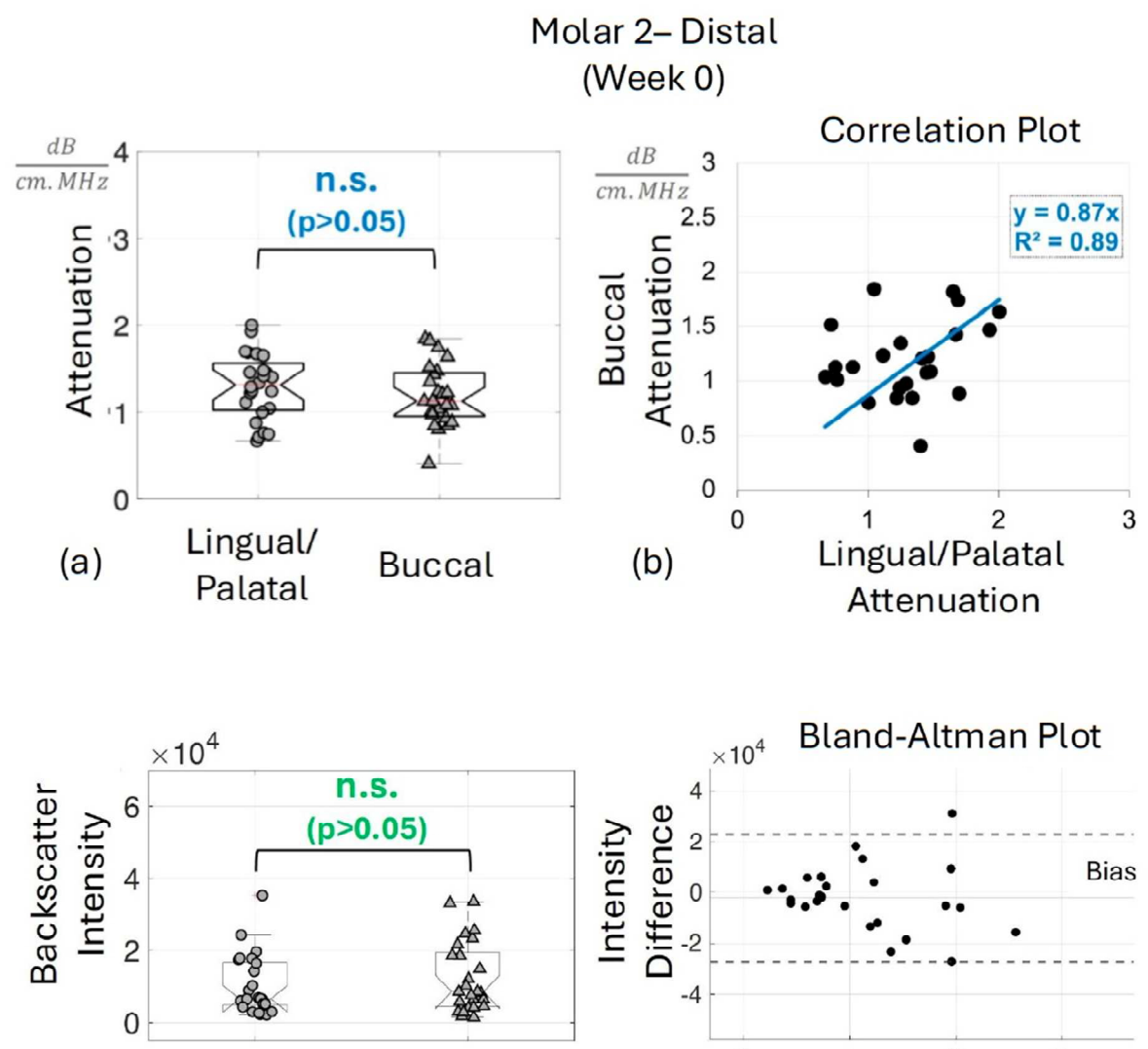

Figure 6: (a) Boxplots comparing ACS between the buccal and lingual-palatal sides at the second molar in healthy gum (n.s.: not significant). (b) Linear correlation of ACS between these sites. (c) Boxplots comparing BSI between buccal and lingual-palatal sides. (d) Bland-Altman plot comparing BSI between these sides.

For BSI, the fitted slope was 0.75 with the confidence interval of 0.46 - 1.28. Given the greater variability observed in BSI compared with ACS, agreement between lingual/palatal and buccal BSI measurements was further assessed using a Bland-Altman plot, as shown in Figure 6 (d). The bias line (solid line) showing average difference being near zero indicates that the two sides do not show a systematic difference in their mean BSIs. Most paired differences fall within the 95% limits of agreement with no obvious trend observed. However, the variability of differences is substantial between individual pairs.

## V. Conclusion

This study presents the first investigation of QUS (attenuation) for longitudinal characterization of pathological conditions in periodontal tissues. Here, we focused on inflammation monitoring as well as providing insights into buccal versus lingual/palatal sides using QUS signatures. Our results demonstrated that inflammation inoculation was associated with decreased ACS. Also, we showed that buccal versus lingual/palatal periodontal tissues surrounding a specific tooth at two sides were not statistically distinct from attenuation and intensity assessments. This study highlights the promising potential of QUS to complement current clinical standards of care in dentistry. Future work will focus on multiparametric QUS for periodontal disease assessment as well as translating QUS techniques to clinical studies.

## Acknowledgment

This work was supported by the funding from the National Institutes of Health, award numbers R21DE029005 (O.D.K and H-L.C.) and F32DE034986 (D.P.). The authors acknowledge Dr. Kerby Shedden, Professor of Statistics at the University of Michigan for valuable statistical consultations.